\documentclass[10pt,twocolumn,showkeys,superscriptaddress,amsmath,amssymb,nofootinbib]{revtex4-2}

\usepackage{graphicx}
\usepackage{bm}
\usepackage{color}
\usepackage{booktabs}
\usepackage{silence}
\usepackage{hyperref}
\usepackage{comment}

\hypersetup{
    colorlinks=true,
    linkcolor=blue,
    filecolor=magenta,      
    urlcolor=blue,
    citecolor=blue,
}

\begin{document}

\title{Discriminating baryonium and final-state-interaction interpretations of $X(2356)$}

\author{Bing-Dong Wan}
\email{wanbd@lnnu.edu.cn}
\affiliation{Department of Physics, Liaoning Normal University, Dalian 116029, China}
\affiliation{Center for Theoretical and Experimental High Energy Physics, Liaoning Normal University, Dalian 116029, China}

\author{Sheng-Qi Zhang}
\email{shqzhang@pku.edu.cn}
\affiliation{Center for High Energy Physics, Peking University, Beijing 100871, China}

\begin{abstract}
We revisit the $X(2356)$ enhancement observed by BESIII in $e^+e^-\to\Lambda\bar\Lambda\eta$ from the viewpoint of near-threshold dynamics. Motivated by the possibility of light $\Lambda\bar\Lambda$ baryonium, we compare this interpretation with final-state-interaction (FSI) explanations within a minimal FSI-dressed pole framework. Fits of the pure-FSI, baryonium-like-pole, and mixed pole--continuum scenarios to the digitized BESIII spectrum identify the energy-dependent FSI description as providing the best balance between fit quality and model complexity. The extracted pole positions depend strongly on the amplitude parametrization, demonstrating the sensitivity of the baryonium interpretation to the treatment of the continuum. We discuss how decay channels, spin observables, and partner searches can provide complementary constraints on the dynamics underlying the enhancement.
\end{abstract}


\maketitle

\section{Introduction}

Baryon-antibaryon systems, historically termed ``baryonium'', are a natural arena for testing multiquark dynamics in nonperturbative QCD. In the one-boson-exchange picture, the $G$-parity transformation relates dibaryon interactions to baryon-antibaryon interactions and can turn part of the short-range repulsion in the baryon-baryon sector into attraction in the baryon-antibaryon sector~\cite{Dover:1979zj,Shapiro:1978wi,Cote:1982gr}. This mechanism has long motivated searches for $N\bar N$ baryonium-like states~\cite{Dalkarov:1970qb,Buck:1977rt,Montanet:1980jy,Amsler:1987qqd,Dover:1990kn,BES:2003aic,BES:2005ega,BESIII:2011aa,Datta:2003iy,Zou:2003zn,Zhu:2005ns,BESIII:2013sbm,Zhang:2025qmg} and for their strange and heavy-flavor analogues~\cite{Wan:2021vny,Wan:2019ake,Chen:2016ymy,Wang:2021qmn}.

The $\Lambda\bar\Lambda$ channel is especially useful because the isoscalar $\Lambda$ forbids diagonal one-pion exchange and its weak decay is self-analyzing, allowing access to spin observables~\cite{Nagels:1978sc,Barnes:1996si}. A structure in this channel therefore probes short- and intermediate-range baryon-antibaryon dynamics in a way complementary to the more familiar $p\bar p$ sector.

Recently, the BESIII collaboration reported an enhancement near the $\Lambda\bar{\Lambda}$ threshold in the process $e^+e^- \to \Lambda\bar{\Lambda}\eta$, which we denote as $X(2356)$~\cite{BESIII:2022tvj}. This structure is characterized by quantum numbers $J^{PC}=1^{--}$, a mass of $(2356 \pm 7 \pm 17)$~MeV/$c^2$, and a width of $(304 \pm 28 \pm 54)$~MeV. A previous QCD sum rule study of the $\Lambda\bar{\Lambda}$ system predicted a vector state in the relevant mass region~\cite{Wan:2021vny}, making the baryonium interpretation a motivated possibility. At the same time, baryon-antibaryon threshold enhancements can be generated without introducing a new compact state. The well-known $p\bar p$ enhancement associated with the $X(1835)$ has been interpreted in terms of FSI effects~\cite{Sibirtsev:2005ds}, and the near-threshold behavior of $e^+e^-\to\Lambda\bar\Lambda$ has also been described by $\Lambda\bar\Lambda$ FSI~\cite{Milstein:2023xzw}. Most directly, Haidenbauer and Mei{\ss}ner described the $e^+e^-\to\eta\Lambda\bar\Lambda$ spectrum within the distorted-wave Born approximation using J\"ulich $\Lambda\bar\Lambda$ potentials constrained by $p\bar p\to\Lambda\bar\Lambda$ data, demonstrating that FSI alone can reproduce the near-threshold mass dependence after an overall normalization~\cite{Haidenbauer:2023llf}. Their analysis was primarily designed to establish the role of the $\Lambda\bar\Lambda$ interaction, while a comparative assessment of explicit-pole and pole--continuum amplitudes lay beyond its scope. 

This Letter therefore does not attempt to establish $X(2356)$ as a pure baryonium state from its mass alone. Instead, we ask what observations can discriminate a baryonium-like pole from an enhancement driven mainly by $\Lambda\bar\Lambda$ FSI. To address this question, we first formulate a minimal FSI-dressed pole framework whose limiting cases correspond to a pure FSI enhancement, a baryonium-like pole, and a mixed pole--continuum scenario. We then fit these alternatives to the digitized BESIII spectrum, compare their fit quality and pole structures, and assess what can be distinguished with the present data. Finally, we clarify the relation of the QCD sum-rule prediction to the bare-state parameter and dressed pole, and discuss how decay channels, partner searches, and polarization observables can further test the baryonium interpretation.

\section{Near-threshold dynamics}

The main limitation of a purely spectroscopic discussion is that a near-threshold enhancement is not equivalent to a new state. In baryon-antibaryon systems the production amplitude can be strongly modified by the interaction between the outgoing baryon and antibaryon. This point is familiar from the $p\bar p$ threshold enhancement in $J/\psi$ decays~\cite{Sibirtsev:2005ds}, from the description of $e^+e^-\to\Lambda\bar\Lambda$ close to threshold in terms of $\Lambda\bar\Lambda$ FSI~\cite{Milstein:2023xzw}, and from the dynamical treatment of $e^+e^-\to\eta\Lambda\bar\Lambda$ and $e^+e^-\to\phi\Lambda\bar\Lambda$~\cite{Haidenbauer:2023llf}. Thus any baryonium interpretation of $X(2356)$ must be formulated together with the continuum interaction.

A compact way to express the issue is to factorize the short-distance production kernel and the low-energy $\Lambda\bar\Lambda$ scattering amplitude,
\begin{equation}
\mathcal{M}(s)=\mathcal{M}_{0}(s)\,F_{\Lambda\bar\Lambda}(s),
\label{eq:fsi_factor}
\end{equation}
where $s=M^2_{\Lambda\bar\Lambda}$ and $\mathcal{M}_{0}$ varies slowly over the narrow threshold region. In an effective-range representation one may write, schematically,
\begin{equation}
F_{\Lambda\bar\Lambda}(k)=\frac{1}{-1/a+(r_e/2)k^2-ik},
\label{eq:effective_range}
\end{equation}
with $k$ the relative momentum and $a$ and $r_e$ the complex scattering length and effective range. We take $k>0$ on the upper rim of the physical cut above the $\Lambda\bar\Lambda$ threshold. The physical and second $\Lambda\bar\Lambda$ sheets correspond to ${\rm Im}\,k>0$ and ${\rm Im}\,k<0$, respectively. With $S=1+2ikF_{\Lambda\bar\Lambda}$, absorption requires ${\rm Im}[-1/a+(r_e/2)k^2]\leq0$ for real $k>0$. A pole close to threshold corresponds to a zero of the denominator after analytic continuation, while a non-pole FSI enhancement can still increase the spectrum through a large low-energy amplitude. This form also makes clear why a line-shape fit alone is not always sufficient: both a nearby pole and a strong attractive FSI can enhance the same mass region.

\begin{figure}[t]
\centering
\includegraphics[width=0.48\textwidth]{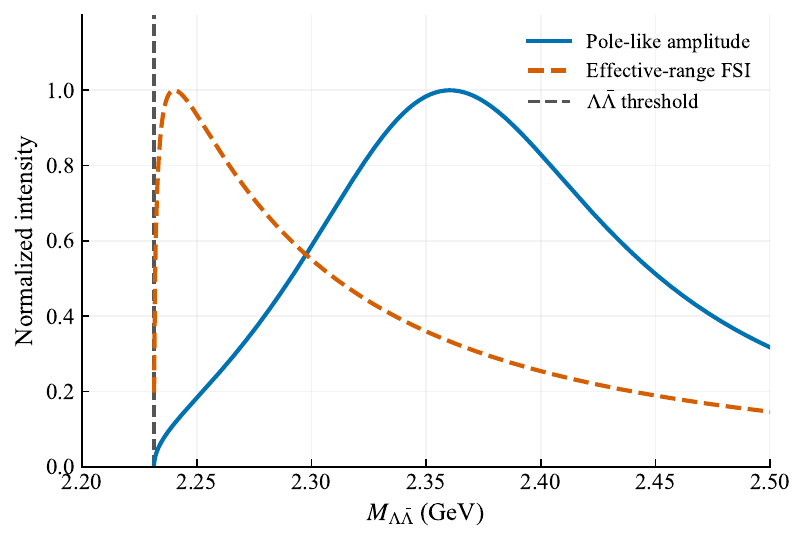}
\caption{Minimal-model comparison of a pole-like near-threshold enhancement and an FSI-only enhancement. The FSI-only curve uses the effective-range parameters in Eq.~(\ref{eq:fsi_parameters}), while the pole-like curve uses the Flatte-inspired parameters in Eq.~(\ref{eq:flatte_parameters}). Both spectra are normalized to their maxima. The vertical line indicates the $\Lambda\bar\Lambda$ threshold.}
\label{fig:lineshape}
\end{figure}

To make this distinction explicit at the level of a minimal model, we evaluate the $\Lambda\bar\Lambda$ spectrum with the two amplitudes used in Fig.~\ref{fig:lineshape}. For the FSI-only curve we use Eq.~(\ref{eq:effective_range}) with
\begin{equation}
a=(-1.0-0.8i)~{\rm fm},\qquad r_e=1.5~{\rm fm},
\label{eq:fsi_parameters}
\end{equation}
and multiply $|F_{\Lambda\bar\Lambda}(k)|^2$ by the two-body phase-space factor $\rho_{\Lambda\bar\Lambda}(s)=2k/\sqrt{s}$. These illustrative values are of order 1~fm and are not fitted to the BESIII data or matched to the interaction in Ref.~\cite{Haidenbauer:2023llf}. With real $r_e$, the negative imaginary part of $a$ satisfies the absorptive condition in the convention of Eq.~(\ref{eq:effective_range}). For the pole-like curve we use a Flatte-inspired denominator with
\begin{equation}
\begin{split}
M_R&=2.355~{\rm GeV},\qquad \Gamma_R=0.12~{\rm GeV},\\
g_{\Lambda\bar\Lambda}&=0.35~{\rm GeV}^2,
\end{split}
\label{eq:flatte_parameters}
\end{equation}
where the imaginary part is taken as $M_R\Gamma_R+g_{\Lambda\bar\Lambda}\rho_{\Lambda\bar\Lambda}(s)$.

If one introduces an explicit baryonium bare pole, the amplitude may be written as
\begin{equation}
\mathcal{M}(s)=\mathcal{M}_{\rm bg}(s)+
\frac{c_{\gamma^*\eta X}\,g_{X\Lambda\bar\Lambda}}
{M_R^2-s-\Sigma_{\Lambda\bar\Lambda}(s)-iM_R\Gamma_{\rm inel}(s)} ,
\label{eq:dressed_pole}
\end{equation}
where the self-energy $\Sigma_{\Lambda\bar\Lambda}$ dresses the bare pole through the near-threshold continuum. Equations~(\ref{eq:fsi_factor})--(\ref{eq:dressed_pole}) organize three scenarios: an FSI amplitude without an explicit bare state, an FSI-dressed baryonium-like pole, and a coherent mixture of continuum and pole contributions.

Figure~\ref{fig:lineshape} illustrates how both FSI and an explicit pole can produce a near-threshold enhancement. For the chosen parameters, the FSI spectrum rises from threshold to a nearby maximum and then decreases smoothly, whereas the pole-like spectrum has a broader maximum at a higher mass. Section~III tests whether the BESIII spectrum distinguishes these scenarios using a common optical $K$-matrix parametrization and identical production and response factors.

\section{Quantitative comparison with the digitized BESIII spectrum}

To determine whether the BESIII spectrum can discriminate among the three scenarios discussed above, we fit the black data points and sideband curve digitized from Fig.~6 of Ref.~\cite{BESIII:2022tvj}. The fit uses the 22 complete bins in $2.24<M_{\Lambda\bar\Lambda}<2.70$~GeV with a common response and background treatment. Four fit variants represent the three physical scenarios: the pure-FSI case is described by constant and energy-dependent continuum amplitudes, while the baryonium-like and mixed cases are represented by an FSI-dressed bare pole and a coherent continuum--pole amplitude, respectively.

With $E=M_{\Lambda\bar\Lambda}-2m_\Lambda$, $k=(M_{\Lambda\bar\Lambda}^2/4-m_\Lambda^2)^{1/2}$, $C(E)=C_0+C_1E/\Lambda_E$, and $d(E)=E_0-E-i\Gamma_{\rm inel}/2$, the three amplitudes are implemented in a one-channel optical $K$-matrix form as
\begin{align}
\mathcal A_{\rm FSI}(E)&=\frac{1}{1-ikC(E)},\nonumber\\
\mathcal D(E)&=d(E)[1-ikC(E)]-ikg^2,\nonumber\\
\mathcal A_{\rm pole}(E)&=\frac{1}{d(E)-ikg^2},\nonumber\\
\mathcal A_{\rm mix}(E)&=\frac{p_c d(E)+g h}{\mathcal D(E)},
\label{eq:fit_amplitude}
\end{align}
where $\Lambda_E=0.3$~GeV, $p_c=\cos\theta$, and $h=\Lambda_E\sin\theta e^{i\phi}$. The constant FSI variant has $C_1=0$, whereas $C_1$ is fitted in the energy-dependent variant. The constraints ${\rm Im}\,C_0\geq0$, $C_1\geq0$, and $\Gamma_{\rm inel}>0$ enforce the absorptive optical convention. All amplitudes are folded with the same phase-space, efficiency, and sideband factors. The fit quality is measured by the minimum Poisson deviance $D$, while AICc accounts for the number of fitted parameters; smaller values indicate a better fit and a better complexity-adjusted description, respectively. A parametric-bootstrap test is used to compare the energy-dependent FSI and mixed amplitudes.

\begin{figure}[t]
\centering
\includegraphics[width=0.48\textwidth]{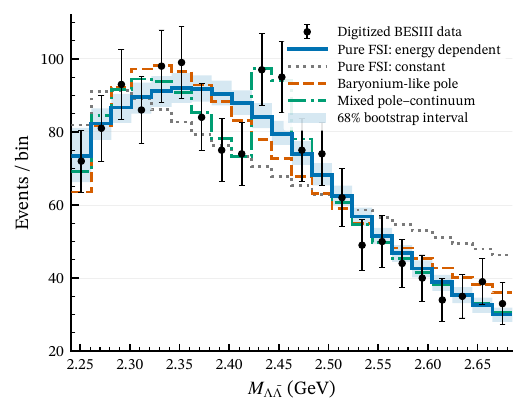}
\caption{Fits to the digitized BESIII combined $M_{\Lambda\bar\Lambda}$ spectrum. All curves use the same response and sideband treatment. The shaded band is the central 68\% pointwise interval from the conditional parametric bootstrap of the energy-dependent FSI fit.}
\label{fig:conditional_fit}
\end{figure}

\begin{table}[t]
\centering
\caption{Fits to the digitized BESIII combined spectrum. Here $n_p$ is the total number of fitted parameters, including the common yield normalization; $D$ is the minimum Poisson deviance; and AICc is the finite-sample corrected Akaike information criterion.}
\label{tab:quantitative_fits}
\scriptsize
\begin{tabular}{lrrr}
\toprule
Model & $n_p$ & $D$ & AICc \\
\midrule
Pure FSI: constant & 3 & 54.64 & 61.98 \\
Pure FSI: energy dependent & 4 & 16.22 & 26.57 \\
Baryonium-like pole & 4 & 24.41 & 34.76 \\
Mixed pole--continuum & 9 & 5.27 & 38.27 \\
\bottomrule
\end{tabular}
\end{table}

The fit results are summarized in Fig.~\ref{fig:conditional_fit} and Table~\ref{tab:quantitative_fits}. Relative to the constant FSI fit, allowing energy dependence adds one parameter and reduces $D$ from 54.64 to 16.22 and AICc from 61.98 to 26.57. The energy-dependent FSI fit also gives lower values of both quantities than the pole-only fit, for which $n_p=4$, $D=24.41$, and ${\rm AICc}=34.76$. It therefore provides the best overall balance between fit quality and the number of parameters. The mixed amplitude achieves a lower deviance, $D=5.27$, but requires $n_p=9$ and gives ${\rm AICc}=38.27$, motivating a statistical assessment of whether the reduction in $D$ supports the additional pole contribution.

We therefore generate 500 Poisson pseudoexperiments under the energy-dependent FSI hypothesis and repeat the model comparison for nine variations of the response, background, fit window, binning, and resolution. The observed $\Delta D=10.95$ corresponds to $p=0.048\pm0.010$, where the uncertainty is the Monte Carlo standard error; the energy-dependent FSI fit retains the lowest AICc in all tested variations, whereas the mixed fit places $\Gamma_{\rm inel}$ at the lower limit of its allowed range. The bootstrap result gives only marginal support for the improvement obtained with the mixed amplitude, while AICc favors the energy-dependent FSI description. Thus, within the models considered, the digitized spectrum does not provide compelling evidence for an additional explicit-pole contribution.

The preference for the energy-dependent FSI amplitude is consistent with Haidenbauer and Mei{\ss}ner~\cite{Haidenbauer:2023llf}, who evaluated the $\Lambda\bar\Lambda$ FSI within the distorted-wave Born approximation using J\"ulich potentials constrained by $p\bar p\to\Lambda\bar\Lambda$ data. After an overall normalization, their $^3S_1$ spectra reproduced the near-threshold BESIII mass dependence, providing a dynamically constrained demonstration that an explicit pole is not required to describe the enhancement. Since their study focused on the predictive role of the FSI interaction, it did not by itself determine whether the same spectrum allows an additional baryonium component. Our analysis complements their calculation by placing the pure-FSI, explicit-pole, and mixed amplitudes within a common fit and comparing their statistical descriptions of the digitized data.

We also examine whether analytic continuation yields a stable pole position. We define the physical $\Lambda\bar\Lambda$ sheet by ${\rm Im}\,k>0$ and the second sheet by ${\rm Im}\,k<0$; in models containing a bare state, the pole is followed continuously from the bare-state limit to the fitted coupling. The energy-dependent FSI amplitude contains a broad second-sheet singularity at $M=2.420-i0.207$~GeV, with 68\% bootstrap intervals ${\rm Re}\,M=[2.385,2.441]$~GeV and ${\rm Im}\,M=[-0.235,-0.180]$~GeV, while the pole continuously connected to the bare state lies at $2.257-i0.182$~GeV in the pole-only fit and $2.431-i0.014$~GeV in the mixed fit. The variation in both the real and imaginary parts indicates that the present spectrum does not determine a model-independent pole mass and width. This single-channel analysis also does not determine the internal composition of a possible state or its couplings to other channels.

\section{Baryonium, QCD sum rules, and mixing}

The fit leaves room for a baryonium contribution, for which the QCD sum-rule prediction provides a complementary motivation. Within the narrow-pole approximation, the vector $\Lambda\bar\Lambda$ result~\cite{Wan:2021vny} estimates the mass scale of the lowest physical state coupled to the chosen local hexaquark current. This mass is distinct from the model-dependent parameter $M_R$ in Eq.~(\ref{eq:dressed_pole}), which precedes the $\Lambda\bar\Lambda$ self-energy correction. The corresponding physical state is associated with a pole of the dressed amplitude. The sum rule thus motivates a possible state near the BESIII mass region, while the line-shape analysis tests whether the spectrum requires its contribution. Quantitative matching would relate the sum-rule mass to the dressed pole with a consistent treatment of the continuum.

A local hexaquark current specifies the quantum numbers of the states to which it couples, but does not uniquely determine their internal composition. The physical state, if it exists, may therefore contain both short-distance and hadronic components and can be written schematically as
\begin{equation}
|X\rangle=a|\Lambda\bar\Lambda\rangle+b|s\bar s\rangle
+c|qq\bar q\bar q\rangle+d|G\rangle+\cdots .
\label{eq:mixing}
\end{equation}
A large $|a|^2$ component implies privileged baryon-antibaryon channels and strong sensitivity to $\Lambda\bar\Lambda$ FSI. A large $|b|^2$ component should be accompanied by ordinary strange-meson modes such as $K\bar K$, $K\bar K^*$, $K^*\bar K^*$, and $\phi\eta$~\cite{Piotrowska:2017rgt}. Hidden-strange tetraquark or gluonic components would instead favor correlated hidden-strange or multimeson channels~\cite{Wan:2025xhf,Liu:2026fsa,Giacosa:2016ypm}. The experimentally useful question is therefore not whether mixing is present, but which component controls the visible decay hierarchy.

\section{Further tests: decay channels, polarization, and partner states}

The decay hierarchy discussed above can be tested by comparing baryon-antibaryon and mesonic channels. Combined amplitude analyses can constrain the channel couplings and examine whether different spectra share a common pole. Together with angular measurements, these comparisons can further constrain the continuum and pole contributions whose separation remains model dependent in Section~III.

Equation~(\ref{eq:fit_amplitude}) gives qualitative expectations for these tests. In the FSI description, the mass dependence is governed by the continuum interaction and the $\Lambda\bar\Lambda$ threshold. In a pole-dominated description, processes coupling to the same state share its pole position, although their observed peak shapes depend on production factors and backgrounds. In the mixed case, interference can enhance or suppress the signal and can produce asymmetric peaks or dips as the relative continuum and pole contributions change. These expectations motivate comparisons across processes within a common amplitude analysis. Since FSI can also generate poles, a shared pole position alone does not distinguish a dynamically generated state from an explicit bare-state contribution.

Searches in other $\Lambda\bar\Lambda$ partial waves can further test the spin dependence of the interaction. Table~\ref{tab:multiplet} lists representative configurations and useful production channels. Whether these configurations support observable partner states depends on the underlying dynamics.

\begin{table}[t]
\centering
\caption{Representative $\Lambda\bar\Lambda$ baryonium configurations including the observed $1^{--}$ channel and possible partners.}
\label{tab:multiplet}
\scriptsize
\begin{tabular}{cp{0.15\columnwidth}p{0.14\columnwidth}p{0.45\columnwidth}}
\toprule
Configuration & Possible $J^{PC}$ & Typical partial wave & Useful probes \\
\midrule
$^1S_0$ & $0^{-+}$ & S & radiative $J/\psi$ decays; $\eta^{(\prime)}B\bar B$ spectra \\
$^3S_1$ & $1^{--}$ & S & $e^+e^-\to\eta\Lambda\bar\Lambda$; time-like form factors \\
$^3P_0$ & $0^{++}$ & P & scalar channels and threshold angular distributions \\
$^1P_1$ & $1^{+-}$ & P & spin-dependent angular distributions and associated production channels \\
$^3P_1$ & $1^{++}$ & P & spin correlations in self-analyzing $\Lambda$ decays \\
$^3P_2$ & $2^{++}$ & P & tensor moments and coupled hyperon-antihyperon channels \\
\bottomrule
\end{tabular}
\end{table}

Companion baryon-antibaryon channels and their experimental roles are summarized in Table~\ref{tab:companion}. The $p\bar p$ sector provides an FSI benchmark~\cite{Sibirtsev:2005ds}. The $\Sigma\bar\Sigma$ and $\Xi\bar\Xi$ channels can test correlations suggested by approximate flavor symmetry and coupled-channel dynamics. If the interactions in these channels are related by approximate flavor symmetry, correlated structures may occur near their respective thresholds, with masses and strengths set by the channel dynamics. Comparing these spectra in similar production processes can test whether a common dynamical description accounts for the structures. Such correlations would constrain the interaction without by themselves distinguishing an explicit state from dynamically generated poles.

\begin{table}[t]
\centering
\caption{Companion baryon-antibaryon channels and their roles in testing the interaction dynamics.}
\label{tab:companion}
\scriptsize
\begin{tabular}{cp{0.35\columnwidth}p{0.42\columnwidth}}
\toprule
Channel & Diagnostic role & Promising measurements \\
\midrule
$p\bar p$ & FSI benchmark; connection with $X(1835)$ & radiative $J/\psi$ decays; form factors \\
$\Lambda\bar\Lambda$ & primary channel for $X(2356)$ & $e^+e^-\to\eta\Lambda\bar\Lambda$ and $e^+e^-\to\Lambda\bar\Lambda$ \\
$\Sigma\bar\Sigma$ & nearby coupled-channel test & scans around hyperon-antihyperon thresholds \\
$\Xi\bar\Xi$ & flavor-SU(3) extension & high-statistics BESIII/Belle II data \\
\bottomrule
\end{tabular}
\end{table}

Polarization observables provide complementary constraints within the $\Lambda\bar\Lambda$ channel. Because the $\Lambda$ decay is self-analyzing, spin correlations can be measured from the decay-product angular distributions~\cite{Barnes:1996si,Aubert:2007uf}. If one common scalar FSI or pole factor multiplies all helicity amplitudes, it cancels from normalized spin observables; a mass enhancement then need not have a corresponding polarization structure. If the continuum and pole terms have different helicity couplings, their interference can instead produce mass-dependent spin correlations. A joint mass--angular analysis can test this possibility. The scalar amplitudes fitted here do not fix those couplings and hence do not predict the sign or size of the polarization effects.

\section{Conclusion}

The $X(2356)$ enhancement lies near the $\Lambda\bar\Lambda$ threshold, where final-state interactions can substantially modify the observed spectrum. We have formulated a minimal FSI-dressed pole framework that accommodates pure-FSI, baryonium-like-pole, and mixed pole--continuum descriptions. Their comparison with the digitized BESIII spectrum favors the energy-dependent FSI amplitude when fit quality and model complexity are considered together. The mixed amplitude improves the fit, but the statistical support for the additional pole is marginal. The extracted pole positions also vary substantially between models, showing that the present spectrum does not determine a model-independent pole mass and width.

We have further clarified that the QCD sum-rule prediction concerns the physical mass scale of a possible state and should be distinguished from the bare-state parameter in the line-shape model. Its coupling to a local hexaquark current does not uniquely determine its internal composition, motivating comparisons of baryon-antibaryon and mesonic decay channels. We have identified conditional expectations for continuum--pole interference and spin correlations, together with partner and companion-channel searches that probe the underlying interaction. These complementary measurements can constrain the pole structure, channel couplings, and decay pattern relevant to the baryonium interpretation.

\begin{acknowledgments}
This work was supported in part by the National Natural Science Foundation of China under Grants 12547114, 12575106, 12325503, and 12147214, and Specific Fund of Fundamental Scientific Research Operating Expenses for Undergraduate Universities in Liaoning Province under Grant No. LJ212410165019. The authors acknowledge the use of Aether (https://aether.aiphys.cn/), an AI agent developed by Peking University, for language editing and manuscript refinement. The authors are fully responsible for the content of this manuscript.
\end{acknowledgments}


\begin{thebibliography}{31}%
\makeatletter
\providecommand \@ifxundefined [1]{%
 \@ifx{#1\undefined}
}%
\providecommand \@ifnum [1]{%
 \ifnum #1\expandafter \@firstoftwo
 \else \expandafter \@secondoftwo
 \fi
}%
\providecommand \@ifx [1]{%
 \ifx #1\expandafter \@firstoftwo
 \else \expandafter \@secondoftwo
 \fi
}%
\providecommand \natexlab [1]{#1}%
\providecommand \enquote  [1]{``#1''}%
\providecommand \bibnamefont  [1]{#1}%
\providecommand \bibfnamefont [1]{#1}%
\providecommand \citenamefont [1]{#1}%
\providecommand \href@noop [0]{\@secondoftwo}%
\providecommand \href [0]{\begingroup \@sanitize@url \@href}%
\providecommand \@href[1]{\@@startlink{#1}\@@href}%
\providecommand \@@href[1]{\endgroup#1\@@endlink}%
\providecommand \@sanitize@url [0]{\catcode `\\12\catcode `\$12\catcode `\&12\catcode `\#12\catcode `\^12\catcode `\_12\catcode `\%12\relax}%
\providecommand \@@startlink[1]{}%
\providecommand \@@endlink[0]{}%
\providecommand \url  [0]{\begingroup\@sanitize@url \@url }%
\providecommand \@url [1]{\endgroup\@href {#1}{\urlprefix }}%
\providecommand \urlprefix  [0]{URL }%
\providecommand \Eprint [0]{\href }%
\providecommand \doibase [0]{https://doi.org/}%
\providecommand \selectlanguage [0]{\@gobble}%
\providecommand \bibinfo  [0]{\@secondoftwo}%
\providecommand \bibfield  [0]{\@secondoftwo}%
\providecommand \translation [1]{[#1]}%
\providecommand \BibitemOpen [0]{}%
\providecommand \bibitemStop [0]{}%
\providecommand \bibitemNoStop [0]{.\EOS\space}%
\providecommand \EOS [0]{\spacefactor3000\relax}%
\providecommand \BibitemShut [1]{\csname bibitem#1\endcsname}%
\let\auto@bib@innerbib\@empty
\bibitem [{\citenamefont {Dover}\ and\ \citenamefont {Richard}(1979)}]{Dover:1979zj}%
  \BibitemOpen
  \bibfield  {author} {\bibinfo {author} {\bibfnamefont {C.~B.}\ \bibnamefont {Dover}}\ and\ \bibinfo {author} {\bibfnamefont {J.~M.}\ \bibnamefont {Richard}},\ }\bibfield  {title} {\bibinfo {title} {{The Interaction of Nucleons With Anti-nucleons. 2. Narrow Mesons Near Threshold: Experiment and Theory}},\ }\href {https://doi.org/10.1016/0003-4916(79)90092-7} {\bibfield  {journal} {\bibinfo  {journal} {Annals Phys.}\ }\textbf {\bibinfo {volume} {121}},\ \bibinfo {pages} {70} (\bibinfo {year} {1979})}\BibitemShut {NoStop}%
\bibitem [{\citenamefont {Shapiro}(1978)}]{Shapiro:1978wi}%
  \BibitemOpen
  \bibfield  {author} {\bibinfo {author} {\bibfnamefont {I.~S.}\ \bibnamefont {Shapiro}},\ }\bibfield  {title} {\bibinfo {title} {{The Physics of Nucleon-antiNucleon Systems}},\ }\href {https://doi.org/10.1016/0370-1573(78)90190-4} {\bibfield  {journal} {\bibinfo  {journal} {Phys. Rept.}\ }\textbf {\bibinfo {volume} {35}},\ \bibinfo {pages} {129} (\bibinfo {year} {1978})}\BibitemShut {NoStop}%
\bibitem [{\citenamefont {Cote}\ \emph {et~al.}(1982)\citenamefont {Cote}, \citenamefont {Lacombe}, \citenamefont {Loiseau}, \citenamefont {Moussallam},\ and\ \citenamefont {Vinh~Mau}}]{Cote:1982gr}%
  \BibitemOpen
  \bibfield  {author} {\bibinfo {author} {\bibfnamefont {J.}~\bibnamefont {Cote}}, \bibinfo {author} {\bibfnamefont {M.}~\bibnamefont {Lacombe}}, \bibinfo {author} {\bibfnamefont {B.}~\bibnamefont {Loiseau}}, \bibinfo {author} {\bibfnamefont {B.}~\bibnamefont {Moussallam}},\ and\ \bibinfo {author} {\bibfnamefont {R.}~\bibnamefont {Vinh~Mau}},\ }\bibfield  {title} {\bibinfo {title} {{On the Nucleon - anti-Nucleon Optical Potential}},\ }\href {https://doi.org/10.1103/PhysRevLett.48.1319} {\bibfield  {journal} {\bibinfo  {journal} {Phys. Rev. Lett.}\ }\textbf {\bibinfo {volume} {48}},\ \bibinfo {pages} {1319} (\bibinfo {year} {1982})}\BibitemShut {NoStop}%
\bibitem [{\citenamefont {Dalkarov}\ \emph {et~al.}(1970)\citenamefont {Dalkarov}, \citenamefont {Mandelzweig},\ and\ \citenamefont {Shapiro}}]{Dalkarov:1970qb}%
  \BibitemOpen
  \bibfield  {author} {\bibinfo {author} {\bibfnamefont {O.~D.}\ \bibnamefont {Dalkarov}}, \bibinfo {author} {\bibfnamefont {V.~B.}\ \bibnamefont {Mandelzweig}},\ and\ \bibinfo {author} {\bibfnamefont {I.~S.}\ \bibnamefont {Shapiro}},\ }\bibfield  {title} {\bibinfo {title} {{On possible quasinuclear nature of heavy meson resonances}},\ }\href {https://doi.org/10.1016/0550-3213(70)90463-3} {\bibfield  {journal} {\bibinfo  {journal} {Nucl. Phys. B}\ }\textbf {\bibinfo {volume} {21}},\ \bibinfo {pages} {88} (\bibinfo {year} {1970})}\BibitemShut {NoStop}%
\bibitem [{\citenamefont {Buck}\ \emph {et~al.}(1979)\citenamefont {Buck}, \citenamefont {Dover},\ and\ \citenamefont {Richard}}]{Buck:1977rt}%
  \BibitemOpen
  \bibfield  {author} {\bibinfo {author} {\bibfnamefont {W.~W.}\ \bibnamefont {Buck}}, \bibinfo {author} {\bibfnamefont {C.~B.}\ \bibnamefont {Dover}},\ and\ \bibinfo {author} {\bibfnamefont {J.~M.}\ \bibnamefont {Richard}},\ }\bibfield  {title} {\bibinfo {title} {{The Interaction of Nucleons with anti-Nucleons. 1. General Features of the anti-N n Spectrum in Potential Models}},\ }\href {https://doi.org/10.1016/0003-4916(79)90091-5} {\bibfield  {journal} {\bibinfo  {journal} {Annals Phys.}\ }\textbf {\bibinfo {volume} {121}},\ \bibinfo {pages} {47} (\bibinfo {year} {1979})}\BibitemShut {NoStop}%
\bibitem [{\citenamefont {Montanet}(1980)}]{Montanet:1980jy}%
  \BibitemOpen
  \bibfield  {author} {\bibinfo {author} {\bibfnamefont {L.}~\bibnamefont {Montanet}},\ }\bibfield  {title} {\bibinfo {title} {{BARYONIUMS: EXPERIMENTAL STATUS}},\ }\href {https://doi.org/10.1016/0370-1573(80)90163-5} {\bibfield  {journal} {\bibinfo  {journal} {Phys. Rept.}\ }\textbf {\bibinfo {volume} {63}},\ \bibinfo {pages} {201} (\bibinfo {year} {1980})}\BibitemShut {NoStop}%
\bibitem [{\citenamefont {Amsler}(1987)}]{Amsler:1987qqd}%
  \BibitemOpen
  \bibfield  {author} {\bibinfo {author} {\bibfnamefont {C.}~\bibnamefont {Amsler}},\ }\bibfield  {title} {\bibinfo {title} {{$\bar{P} P$ Interaction and the Quest for Baryonium}},\ }\href@noop {} {\bibfield  {journal} {\bibinfo  {journal} {Adv. Nucl. Phys.}\ }\textbf {\bibinfo {volume} {18}},\ \bibinfo {pages} {183} (\bibinfo {year} {1987})}\BibitemShut {NoStop}%
\bibitem [{\citenamefont {Dover}\ \emph {et~al.}(1991)\citenamefont {Dover}, \citenamefont {Gutsche},\ and\ \citenamefont {Faessler}}]{Dover:1990kn}%
  \BibitemOpen
  \bibfield  {author} {\bibinfo {author} {\bibfnamefont {C.~B.}\ \bibnamefont {Dover}}, \bibinfo {author} {\bibfnamefont {T.}~\bibnamefont {Gutsche}},\ and\ \bibinfo {author} {\bibfnamefont {A.}~\bibnamefont {Faessler}},\ }\bibfield  {title} {\bibinfo {title} {{The Case for quasinuclear $\bar{N}N$ bound states}},\ }\href {https://doi.org/10.1103/PhysRevC.43.379} {\bibfield  {journal} {\bibinfo  {journal} {Phys. Rev. C}\ }\textbf {\bibinfo {volume} {43}},\ \bibinfo {pages} {379} (\bibinfo {year} {1991})}\BibitemShut {NoStop}%
\bibitem [{\citenamefont {Bai}\ \emph {et~al.}(2003)\citenamefont {Bai} \emph {et~al.}}]{BES:2003aic}%
  \BibitemOpen
  \bibfield  {author} {\bibinfo {author} {\bibfnamefont {J.~Z.}\ \bibnamefont {Bai}} \emph {et~al.} (\bibinfo {collaboration} {BES}),\ }\bibfield  {title} {\bibinfo {title} {{Observation of a near threshold enhancement in the $p \bar{p}$ mass spectrum from radiative $J / \psi \rightarrow \gamma p \bar{p}$ decays}},\ }\href {https://doi.org/10.1103/PhysRevLett.91.022001} {\bibfield  {journal} {\bibinfo  {journal} {Phys. Rev. Lett.}\ }\textbf {\bibinfo {volume} {91}},\ \bibinfo {pages} {022001} (\bibinfo {year} {2003})},\ \Eprint {https://arxiv.org/abs/hep-ex/0303006} {arXiv:hep-ex/0303006} \BibitemShut {NoStop}%
\bibitem [{\citenamefont {Ablikim}\ \emph {et~al.}(2005)\citenamefont {Ablikim} \emph {et~al.}}]{BES:2005ega}%
  \BibitemOpen
  \bibfield  {author} {\bibinfo {author} {\bibfnamefont {M.}~\bibnamefont {Ablikim}} \emph {et~al.} (\bibinfo {collaboration} {BES}),\ }\bibfield  {title} {\bibinfo {title} {{Observation of a resonance X(1835) in $J / \psi \rightarrow \gamma \pi^{+} \pi^{-} \eta^{\prime}$}},\ }\href {https://doi.org/10.1103/PhysRevLett.95.262001} {\bibfield  {journal} {\bibinfo  {journal} {Phys. Rev. Lett.}\ }\textbf {\bibinfo {volume} {95}},\ \bibinfo {pages} {262001} (\bibinfo {year} {2005})},\ \Eprint {https://arxiv.org/abs/hep-ex/0508025} {arXiv:hep-ex/0508025} \BibitemShut {NoStop}%
\bibitem [{\citenamefont {Ablikim}\ \emph {et~al.}(2012)\citenamefont {Ablikim} \emph {et~al.}}]{BESIII:2011aa}%
  \BibitemOpen
  \bibfield  {author} {\bibinfo {author} {\bibfnamefont {M.}~\bibnamefont {Ablikim}} \emph {et~al.} (\bibinfo {collaboration} {BESIII}),\ }\bibfield  {title} {\bibinfo {title} {{Spin-Parity Analysis of $p\bar{p}$ Mass Threshold Structure in $J/\psi$ and $\psi^\prime$ Radiative Decays}},\ }\href {https://doi.org/10.1103/PhysRevLett.108.112003} {\bibfield  {journal} {\bibinfo  {journal} {Phys. Rev. Lett.}\ }\textbf {\bibinfo {volume} {108}},\ \bibinfo {pages} {112003} (\bibinfo {year} {2012})},\ \Eprint {https://arxiv.org/abs/1112.0942} {arXiv:1112.0942 [hep-ex]} \BibitemShut {NoStop}%
\bibitem [{\citenamefont {Datta}\ and\ \citenamefont {O'Donnell}(2003)}]{Datta:2003iy}%
  \BibitemOpen
  \bibfield  {author} {\bibinfo {author} {\bibfnamefont {A.}~\bibnamefont {Datta}}\ and\ \bibinfo {author} {\bibfnamefont {P.~J.}\ \bibnamefont {O'Donnell}},\ }\bibfield  {title} {\bibinfo {title} {{A New state of baryonium}},\ }\href {https://doi.org/10.1016/j.physletb.2003.06.050} {\bibfield  {journal} {\bibinfo  {journal} {Phys. Lett. B}\ }\textbf {\bibinfo {volume} {567}},\ \bibinfo {pages} {273} (\bibinfo {year} {2003})},\ \Eprint {https://arxiv.org/abs/hep-ph/0306097} {arXiv:hep-ph/0306097} \BibitemShut {NoStop}%
\bibitem [{\citenamefont {Zou}\ and\ \citenamefont {Chiang}(2004)}]{Zou:2003zn}%
  \BibitemOpen
  \bibfield  {author} {\bibinfo {author} {\bibfnamefont {B.~S.}\ \bibnamefont {Zou}}\ and\ \bibinfo {author} {\bibfnamefont {H.~C.}\ \bibnamefont {Chiang}},\ }\bibfield  {title} {\bibinfo {title} {{One pion exchange final state interaction and the $p \bar{p}$ near threshold enhancement in $J / \psi \rightarrow \gamma p \bar{p}$ decays}},\ }\href {https://doi.org/10.1103/PhysRevD.69.034004} {\bibfield  {journal} {\bibinfo  {journal} {Phys. Rev. D}\ }\textbf {\bibinfo {volume} {69}},\ \bibinfo {pages} {034004} (\bibinfo {year} {2004})},\ \Eprint {https://arxiv.org/abs/hep-ph/0309273} {arXiv:hep-ph/0309273} \BibitemShut {NoStop}%
\bibitem [{\citenamefont {Zhu}\ and\ \citenamefont {Gao}(2006)}]{Zhu:2005ns}%
  \BibitemOpen
  \bibfield  {author} {\bibinfo {author} {\bibfnamefont {S.-L.}\ \bibnamefont {Zhu}}\ and\ \bibinfo {author} {\bibfnamefont {C.-S.}\ \bibnamefont {Gao}},\ }\bibfield  {title} {\bibinfo {title} {{$X(1835)$: A Possible baryonium?}},\ }\href {https://doi.org/10.1088/0253-6102/46/2/021} {\bibfield  {journal} {\bibinfo  {journal} {Commun. Theor. Phys.}\ }\textbf {\bibinfo {volume} {46}},\ \bibinfo {pages} {291} (\bibinfo {year} {2006})},\ \Eprint {https://arxiv.org/abs/hep-ph/0507050} {arXiv:hep-ph/0507050} \BibitemShut {NoStop}%
\bibitem [{\citenamefont {Ablikim}\ \emph {et~al.}(2013{\natexlab{b}})\citenamefont {Ablikim} \emph {et~al.}}]{BESIII:2013sbm}%
  \BibitemOpen
  \bibfield  {author} {\bibinfo {author} {\bibfnamefont {M.}~\bibnamefont {Ablikim}} \emph {et~al.} (\bibinfo {collaboration} {BESIII}),\ }\bibfield  {title} {\bibinfo {title} {{Observation of a structure at $1.84 \,\mathrm{GeV} / c^2$ in the $3\left(\pi^{+} \pi^{-}\right)$mass spectrum in $J / \psi \rightarrow \gamma 3\left(\pi^{+} \pi^{-}\right)$decays}},\ }\href {https://doi.org/10.1103/PhysRevD.88.091502} {\bibfield  {journal} {\bibinfo  {journal} {Phys. Rev. D}\ }\textbf {\bibinfo {volume} {88}},\ \bibinfo {pages} {091502} (\bibinfo {year} {2013}{\natexlab{b}})},\ \Eprint {https://arxiv.org/abs/1305.5333} {arXiv:1305.5333 [hep-ex]} \BibitemShut {NoStop}%
\bibitem [{\citenamefont {Zhang}\ and\ \citenamefont {Qiao}(2026)}]{Zhang:2025qmg}%
  \BibitemOpen
  \bibfield  {author} {\bibinfo {author} {\bibfnamefont {S.-Q.}\ \bibnamefont {Zhang}}\ and\ \bibinfo {author} {\bibfnamefont {C.-F.}\ \bibnamefont {Qiao}},\ }\bibfield  {title} {\bibinfo {title} {{Baryons and baryoniums in the perspective of QCD sum rules}},\ }\href {https://doi.org/10.1007/s43673-026-00192-y} {\bibfield  {journal} {\bibinfo  {journal} {AAPPS Bull.}\ }\textbf {\bibinfo {volume} {36}},\ \bibinfo {pages} {12} (\bibinfo {year} {2026})},\ \Eprint {https://arxiv.org/abs/2512.24706} {arXiv:2512.24706 [hep-ph]} \BibitemShut {NoStop}%
\bibitem [{\citenamefont {Wan}\ \emph {et~al.}(2022)\citenamefont {Wan}, \citenamefont {Zhang},\ and\ \citenamefont {Qiao}}]{Wan:2021vny}%
  \BibitemOpen
  \bibfield  {author} {\bibinfo {author} {\bibfnamefont {B.-D.}\ \bibnamefont {Wan}}, \bibinfo {author} {\bibfnamefont {S.-Q.}\ \bibnamefont {Zhang}},\ and\ \bibinfo {author} {\bibfnamefont {C.-F.}\ \bibnamefont {Qiao}},\ }\bibfield  {title} {\bibinfo {title} {{Light baryonium spectrum}},\ }\href {https://doi.org/10.1103/PhysRevD.105.014016} {\bibfield  {journal} {\bibinfo  {journal} {Phys. Rev. D}\ }\textbf {\bibinfo {volume} {105}},\ \bibinfo {pages} {014016} (\bibinfo {year} {2022})},\ \Eprint {https://arxiv.org/abs/2109.07130} {arXiv:2109.07130 [hep-ph]} \BibitemShut {NoStop}%
\bibitem [{\citenamefont {Wan}\ \emph {et~al.}(2020)\citenamefont {Wan}, \citenamefont {Tang},\ and\ \citenamefont {Qiao}}]{Wan:2019ake}%
  \BibitemOpen
  \bibfield  {author} {\bibinfo {author} {\bibfnamefont {B.-D.}\ \bibnamefont {Wan}}, \bibinfo {author} {\bibfnamefont {L.}~\bibnamefont {Tang}},\ and\ \bibinfo {author} {\bibfnamefont {C.-F.}\ \bibnamefont {Qiao}},\ }\bibfield  {title} {\bibinfo {title} {{Hidden-bottom and -charm hexaquark states in QCD sum rules}},\ }\href {https://doi.org/10.1140/epjc/s10052-020-7701-8} {\bibfield  {journal} {\bibinfo  {journal} {Eur. Phys. J. C}\ }\textbf {\bibinfo {volume} {80}},\ \bibinfo {pages} {121} (\bibinfo {year} {2020})},\ \Eprint {https://arxiv.org/abs/1912.12046} {arXiv:1912.12046 [hep-ph]} \BibitemShut {NoStop}%
\bibitem [{\citenamefont {Chen}\ \emph {et~al.}(2016)\citenamefont {Chen}, \citenamefont {Zhou}, \citenamefont {Chen}, \citenamefont {Liu},\ and\ \citenamefont {Zhu}}]{Chen:2016ymy}%
  \BibitemOpen
  \bibfield  {author} {\bibinfo {author} {\bibfnamefont {H.-X.}\ \bibnamefont {Chen}}, \bibinfo {author} {\bibfnamefont {D.}~\bibnamefont {Zhou}}, \bibinfo {author} {\bibfnamefont {W.}~\bibnamefont {Chen}}, \bibinfo {author} {\bibfnamefont {X.}~\bibnamefont {Liu}},\ and\ \bibinfo {author} {\bibfnamefont {S.-L.}\ \bibnamefont {Zhu}},\ }\bibfield  {title} {\bibinfo {title} {{Searching for hidden-charm baryonium signals in QCD sum rules}},\ }\href {https://doi.org/10.1140/epjc/s10052-016-4459-0} {\bibfield  {journal} {\bibinfo  {journal} {Eur. Phys. J. C}\ }\textbf {\bibinfo {volume} {76}},\ \bibinfo {pages} {602} (\bibinfo {year} {2016})},\ \Eprint {https://arxiv.org/abs/1605.07453} {arXiv:1605.07453 [hep-ph]} \BibitemShut {NoStop}%
\bibitem [{\citenamefont {Wang}\ \emph {et~al.}(2021)\citenamefont {Wang}, \citenamefont {Wang},\ and\ \citenamefont {Yu}}]{Wang:2021qmn}%
  \BibitemOpen
  \bibfield  {author} {\bibinfo {author} {\bibfnamefont {X.-W.}\ \bibnamefont {Wang}}, \bibinfo {author} {\bibfnamefont {Z.-G.}\ \bibnamefont {Wang}},\ and\ \bibinfo {author} {\bibfnamefont {G.-l.}\ \bibnamefont {Yu}},\ }\bibfield  {title} {\bibinfo {title} {{Study of $\Lambda _c\Lambda _c$ dibaryon and $\Lambda _c{\bar{\Lambda }}_c$ baryonium states via QCD sum rules}},\ }\href {https://doi.org/10.1140/epja/s10050-021-00576-8} {\bibfield  {journal} {\bibinfo  {journal} {Eur. Phys. J. A}\ }\textbf {\bibinfo {volume} {57}},\ \bibinfo {pages} {275} (\bibinfo {year} {2021})},\ \Eprint {https://arxiv.org/abs/2107.04751} {arXiv:2107.04751 [hep-ph]} \BibitemShut {NoStop}%
\bibitem [{\citenamefont {Nagels}\ \emph {et~al.}(1979)\citenamefont {Nagels}, \citenamefont {Rijken},\ and\ \citenamefont {de~Swart}}]{Nagels:1978sc}%
  \BibitemOpen
  \bibfield  {author} {\bibinfo {author} {\bibfnamefont {M.~M.}\ \bibnamefont {Nagels}}, \bibinfo {author} {\bibfnamefont {T.~A.}\ \bibnamefont {Rijken}},\ and\ \bibinfo {author} {\bibfnamefont {J.~J.}\ \bibnamefont {de~Swart}},\ }\bibfield  {title} {\bibinfo {title} {{Baryon Baryon Scattering in a One Boson Exchange Potential Approach. 3. A Nucleon-Nucleon and Hyperon - Nucleon Analysis Including Contributions of a Nonet of Scalar Mesons}},\ }\href {https://doi.org/10.1103/PhysRevD.20.1633} {\bibfield  {journal} {\bibinfo  {journal} {Phys. Rev. D}\ }\textbf {\bibinfo {volume} {20}},\ \bibinfo {pages} {1633} (\bibinfo {year} {1979})}\BibitemShut {NoStop}%
\bibitem [{\citenamefont {Barnes}\ \emph {et~al.}(1996)\citenamefont {Barnes} \emph {et~al.}}]{Barnes:1996si}%
  \BibitemOpen
  \bibfield  {author} {\bibinfo {author} {\bibfnamefont {P.~D.}\ \bibnamefont {Barnes}} \emph {et~al.},\ }\bibfield  {title} {\bibinfo {title} {{Observables in high statistics measurements of the reaction $\bar{p}p\to \bar{\Lambda}\Lambda$}},\ }\href {https://doi.org/10.1103/PhysRevC.54.1877} {\bibfield  {journal} {\bibinfo  {journal} {Phys. Rev. C}\ }\textbf {\bibinfo {volume} {54}},\ \bibinfo {pages} {1877} (\bibinfo {year} {1996})}\BibitemShut {NoStop}%
\bibitem [{\citenamefont {Ablikim}\ \emph {et~al.}(2023)\citenamefont {Ablikim} \emph {et~al.}}]{BESIII:2022tvj}%
  \BibitemOpen
  \bibfield  {author} {\bibinfo {author} {\bibfnamefont {M.}~\bibnamefont {Ablikim}} \emph {et~al.} (\bibinfo {collaboration} {BESIII}),\ }\bibfield  {title} {\bibinfo {title} {{Measurement of $e^+e^-\rightarrow\Lambda\bar{\Lambda}\eta$ from 3.5106 to 4.6988 GeV and study of $\Lambda\bar{\Lambda}$ mass threshold enhancement}},\ }\href {https://doi.org/10.1103/PhysRevD.107.112001} {\bibfield  {journal} {\bibinfo  {journal} {Phys. Rev. D}\ }\textbf {\bibinfo {volume} {107}},\ \bibinfo {pages} {112001} (\bibinfo {year} {2023})},\ \Eprint {https://arxiv.org/abs/2211.10755} {arXiv:2211.10755 [hep-ex]} \BibitemShut {NoStop}%
\bibitem [{\citenamefont {Sibirtsev}\ \emph {et~al.}(2005)\citenamefont {Sibirtsev}, \citenamefont {Haidenbauer}, \citenamefont {Krewald}, \citenamefont {Meissner},\ and\ \citenamefont {Thomas}}]{Sibirtsev:2005ds}%
  \BibitemOpen
  \bibfield  {author} {\bibinfo {author} {\bibfnamefont {A.}~\bibnamefont {Sibirtsev}}, \bibinfo {author} {\bibfnamefont {J.}~\bibnamefont {Haidenbauer}}, \bibinfo {author} {\bibfnamefont {S.}~\bibnamefont {Krewald}}, \bibinfo {author} {\bibfnamefont {U.-G.}\ \bibnamefont {Meissner}},\ and\ \bibinfo {author} {\bibfnamefont {A.~W.}\ \bibnamefont {Thomas}},\ }\bibfield  {title} {\bibinfo {title} {{Near threshold enhancement of the $p\bar p$ mass spectrum in $J/\psi$ decay}},\ }\href {https://doi.org/10.1103/PhysRevD.71.054010} {\bibfield  {journal} {\bibinfo  {journal} {Phys. Rev. D}\ }\textbf {\bibinfo {volume} {71}},\ \bibinfo {pages} {054010} (\bibinfo {year} {2005})}\BibitemShut {NoStop}%
\bibitem [{\citenamefont {Milstein}\ and\ \citenamefont {Salnikov}(2023)}]{Milstein:2023xzw}%
  \BibitemOpen
  \bibfield  {author} {\bibinfo {author} {\bibfnamefont {A.~I.}\ \bibnamefont {Milstein}}\ and\ \bibinfo {author} {\bibfnamefont {S.~G.}\ \bibnamefont {Salnikov}},\ }\bibfield  {title} {\bibinfo {title} {{Natural explanation of recent results on $e^+e^- \to \Lambda\bar\Lambda$}},\ }\href {https://doi.org/10.1134/S0021364023601471} {\bibfield  {journal} {\bibinfo  {journal} {JETP Lett.}\ }\textbf {\bibinfo {volume} {117}},\ \bibinfo {pages} {905} (\bibinfo {year} {2023})}\BibitemShut {NoStop}%
\bibitem [{\citenamefont {Haidenbauer}\ and\ \citenamefont {Meissner}(2023)}]{Haidenbauer:2023llf}%
  \BibitemOpen
  \bibfield  {author} {\bibinfo {author} {\bibfnamefont {J.}~\bibnamefont {Haidenbauer}}\ and\ \bibinfo {author} {\bibfnamefont {U.-G.}\ \bibnamefont {Mei{\ss}ner}},\ }\bibfield  {title} {\bibinfo {title} {{$\Lambda\bar\Lambda$ final-state interaction in the reactions $e^+e^- \to \phi\Lambda\bar\Lambda$ and $e^+e^- \to \eta\Lambda\bar\Lambda$}},\ }\href {https://doi.org/10.1140/epja/s10050-023-01017-4} {\bibfield  {journal} {\bibinfo  {journal} {Eur. Phys. J. A}\ }\textbf {\bibinfo {volume} {59}},\ \bibinfo {pages} {136} (\bibinfo {year} {2023})},\ \Eprint {https://arxiv.org/abs/2303.05128} {arXiv:2303.05128 [nucl-th]} \BibitemShut {NoStop}%
\bibitem [{\citenamefont {Piotrowska}\ and\ \citenamefont {Giacosa}(2017)}]{Piotrowska:2017rgt}%
  \BibitemOpen
  \bibfield  {author} {\bibinfo {author} {\bibfnamefont {M.}~\bibnamefont {Piotrowska}}\ and\ \bibinfo {author} {\bibfnamefont {F.}~\bibnamefont {Giacosa}},\ }\bibfield  {title} {\bibinfo {title} {{Strong decays of excited vector mesons}},\ }\href@noop {} {\bibfield  {journal} {\bibinfo  {journal} {arXiv}\ } (\bibinfo {year} {2017})},\ \Eprint {https://arxiv.org/abs/1708.03175} {arXiv:1708.03175 [hep-ph]} \BibitemShut {NoStop}%
\bibitem [{\citenamefont {Wan}\ and\ \citenamefont {Yang}(2026)}]{Wan:2025xhf}%
  \BibitemOpen
  \bibfield  {author} {\bibinfo {author} {\bibfnamefont {B.-D.}\ \bibnamefont {Wan}}\ and\ \bibinfo {author} {\bibfnamefont {J.-C.}\ \bibnamefont {Yang}},\ }\bibfield  {title} {\bibinfo {title} {{Fully strange tetraquark states via QCD sum rules*}},\ }\href {https://doi.org/10.1088/1674-1137/ae2fc8} {\bibfield  {journal} {\bibinfo  {journal} {Chin. Phys. C}\ }\textbf {\bibinfo {volume} {50}},\ \bibinfo {pages} {043104} (\bibinfo {year} {2026})},\ \Eprint {https://arxiv.org/abs/2507.11874} {arXiv:2507.11874 [hep-ph]} \BibitemShut {NoStop}%
\bibitem [{\citenamefont {Liu}\ \emph {et~al.}(2026)\citenamefont {Liu}, \citenamefont {Zhong},\ and\ \citenamefont {Zhao}}]{Liu:2026fsa}%
  \BibitemOpen
  \bibfield  {author} {\bibinfo {author} {\bibfnamefont {F.-X.}\ \bibnamefont {Liu}}, \bibinfo {author} {\bibfnamefont {X.-H.}\ \bibnamefont {Zhong}},\ and\ \bibinfo {author} {\bibfnamefont {Q.}~\bibnamefont {Zhao}},\ }\bibfield  {title} {\bibinfo {title} {{Fully-strange tetraquarks: fall-apart decays and experimental candidates}},\ }\href@noop {} {\bibfield  {journal} {\bibinfo  {journal} {arXiv}\ } (\bibinfo {year} {2026})},\ \Eprint {https://arxiv.org/abs/2601.03614} {arXiv:2601.03614 [hep-ph]} \BibitemShut {NoStop}%
\bibitem [{\citenamefont {Giacosa}\ \emph {et~al.}(2017)\citenamefont {Giacosa}, \citenamefont {Sammet},\ and\ \citenamefont {Janowski}}]{Giacosa:2016ypm}%
  \BibitemOpen
  \bibfield  {author} {\bibinfo {author} {\bibfnamefont {F.}~\bibnamefont {Giacosa}}, \bibinfo {author} {\bibfnamefont {J.}~\bibnamefont {Sammet}},\ and\ \bibinfo {author} {\bibfnamefont {S.}~\bibnamefont {Janowski}},\ }\bibfield  {title} {\bibinfo {title} {{Decays of the vector glueball}},\ }\href {https://doi.org/10.1103/PhysRevD.95.114004} {\bibfield  {journal} {\bibinfo  {journal} {Phys. Rev. D}\ }\textbf {\bibinfo {volume} {95}},\ \bibinfo {pages} {114004} (\bibinfo {year} {2017})},\ \Eprint {https://arxiv.org/abs/1607.03640} {arXiv:1607.03640 [hep-ph]} \BibitemShut {NoStop}%
\bibitem [{\citenamefont {Aubert}\ \emph {et~al.}(2007)\citenamefont {Aubert} \emph {et~al.}}]{Aubert:2007uf}%
  \BibitemOpen
  \bibfield  {author} {\bibinfo {author} {\bibfnamefont {B.}~\bibnamefont {Aubert}} \emph {et~al.} (\bibinfo {collaboration} {BaBar}),\ }\bibfield  {title} {\bibinfo {title} {{Study of $e^+e^- \to \Lambda \bar{\Lambda}$, $\Lambda \bar{\Sigma}^{0}$, $\Sigma^{0}\bar{\Sigma}^{0}$ using initial state radiation with BABAR}},\ }\href {https://doi.org/10.1103/PhysRevD.76.092006} {\bibfield  {journal} {\bibinfo  {journal} {Phys. Rev. D}\ }\textbf {\bibinfo {volume} {76}},\ \bibinfo {pages} {092006} (\bibinfo {year} {2007})},\ \Eprint {https://arxiv.org/abs/0709.1988} {arXiv:0709.1988 [hep-ex]} \BibitemShut {NoStop}%
\end{thebibliography}
\end{document}